\documentclass[twocolumn,noshowpacs,preprintnumbers,amsmath,amssymb]{revtex4}

\usepackage{verbatim}

\usepackage{graphicx}
\usepackage{dcolumn}
\usepackage{bm}
\usepackage{wrapfig}

\begin{document}

\title{Carbon supported CdSe nanocrystals}
\author{Beatriz H. Juarez}
\email{hernande@chemie.uni-hamburg.de}
\author{Michaela Meyns}
\author{Alina Chanaewa}
\author{Yuxue Cai}
\author{Christian Klinke}
\author{Horst Weller}
\affiliation{Institute of Physical Chemistry, University of Hamburg, D - 20146 Hamburg, Germany}

\begin{abstract} 

Insights to the mechanism of CdSe nanoparticle attachment to carbon nanotubes following the hot injection method are discussed. It was observed that the presence of water improves the nanotube coverage while Cl containing media are responsible for the shape transformation of the nanoparticles and further attachment to the carbon lattice. The experiments also show that the mechanism taking place involves the right balance of several factors, namely, low passivated nanoparticle surface, particles with well-defined crystallographic facets, and interaction with an organics-free sp$^{2}$ carbon lattice. Furthermore, this procedure can be extended to cover graphene by quantum dots.

\end{abstract}

\maketitle

The extraordinary properties of nanosized materials have boosted intensive work in the development of several synthetic methodologies in the last decades. The so called hot injection method~\cite{C01} is, among them, a versatile and relatively cheap means that nowadays provides a high degree of perfection, reproducibility and control of semiconductor~\cite{C02,C03}, and magnetic~\cite{C01,C04} nanocrystals. This colloidal synthetic route is based on the reaction of highly reactive species in media where surfactants play an essential role in the final shape, size, and stability of the desired nanomaterial. However, small variations in monomer concentrations, temperatures, time, or presence of impurities may drive the reaction into different regimes. As a consequence, diverse sizes, shapes, and compositions of nanoparticles can be obtained. A. J. Houtepen et al.~\cite{C05} describes the role of acetate in the synthesis of PbSe nanoparticles. Traces of acetate can be responsible for a great variety of nanoparticles shapes. It is also known that small traces of phosphonic acids can influence decisively the mechanism, the yield, and the shape of the nanoparticles~\cite{C06,C07,C08}. Recently, the hot injection method was also followed to produce composites of CdSe semiconductor nanoparticles and carbon nanotubes (CdSe-CNTs)~\cite{C09,C10,C11}. It was observed that in the presence of CNTs, CdSe nanorods evolved into pyramidal-like nanoparticles that connected to CNTs by the wurtzite (001) facets~\cite{C09,C10,C11,C12}. Furthermore, rods and pyramidal-like nanoparticles showed different interaction with CNTs, the latter showing a clear tendency to attach to the carbon lattice. Unlike heterogeneous growth on a seed surface, the reaction of CdSe nanorods in the presence of carbon nanotubes proceeds independently of the carbon lattice during their nucleation and growth. The proposed mechanism so far~\cite{C09} suggested that the shape transformation of the nanoparticles as a result of nanotube-nanoparticle interaction and a decisive role of octadecylphosphonic acid (ODPA). ODPA is used as complexing agent of the cadmium source and acts as capping ligand of the nanoparticles at Cd sites~\cite{C13}.

In this letter, we address the influence of the CNTs dispersant (solvent) and water on the shape of the nanoparticles and the requirements for further attachment to the carbon lattice. The experiments reported here show that the final shape of the nanoparticles (pyramidal-like) is related to the injection of DCE (1,2-dichloroethane), the carbon-material dispersant. We ascertain that the complex formed by Cd and ODPA is influenced by DCE, triggering a reaction where rod-like nanoparticles change their shape and attach to sp$^{2}$ carbon lattices. The influence of water, the possible formation of HCl as well as the requirements for the attachment of the nanoparticles are also discussed. Their influence has also been studied in some other carbon allotropic systems like diamond powder, glassy carbon, and few-layer graphene exfoliated from highly oriented pyrolitic graphite (HOPG).

\begin{figure}[!h]
\begin{center}
\includegraphics[width=0.45\textwidth]{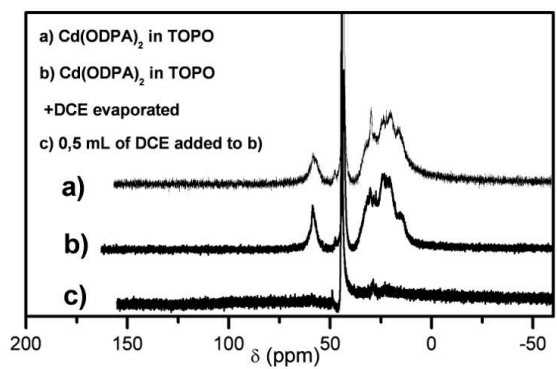}
\caption{\it $^{31}$P $\{$1H$\}$ NMR chemical shifts of a (a) Cd-ODPA mixture in TOPO (b) after addition and elimination under vacuum of 2~mL of DCE and (c) after the addition of 0.5~mL of DCE at RT.}
\end{center}
\end{figure}

The synthesis of CdSe-CNT composites is based on the complexation of CdO by ODPA in trioctylphospinoxide, (TOPO) as solvent, followed by the injection of CNT in DCE and the subsequent removal of the solvent under vacuum (procedure A in the supporting information). After this step, the injection of Se dissolved in trioctylphosphine (TOP) promotes the formation of CdSe nuclei that grow along the c-axis of the wurtzite crystal. We observed that the step of removal of DCE is crucial, as the non-efficient removal of DCE (i.e., as a result of low vacuum power) may delay or even suppress the nucleation of CdSe nanoparticles upon Se injection. Thus, small traces of DCE in solution or adsorbed on the CNTs may influence the formation of the nanoparticles and/or further evolution of their shape. In order to address the effect of DCE on the Cd-ODPA complex, we carried out $^{31}$P-Nuclear Magnetic Resonance (NMR) analyses. Fig.~1a shows the $^{31}$P-NMR chemical shifts of a Cd-ODPA mixture in TOPO (see supporting information for further details). The spectrum contains 3 prominent features. A peak at 58~ppm, a set of multiplets around 25~ppm and a sharp signal at 48~ppm.

A similar spectrum is observed (Fig.~1b) when 2~mL of DCE are injected and removed under vacuum. No appreciable changes (at least, no detectable by NMR means) can be identified if the DCE is quickly removed. For comparison, the spectrum of this mixture in the presence of 0.5~mL of DCE added at room temperature and not evaporated is shown in c). The presence of DCE removes the signals centered around 25~ppm previously assigned to the complex Cd-ODPA~\cite{C06}. The same happens with the signal around 58~ppm, which may originate as a result of a Cd(ODPA)-TOPO hydrogen bonding interaction. The intensity of the peak related to TOPO (around 48~ppm) slightly increases and shifts, as a result of the new chemical environment upon the addition of DCE. 

\begin{figure}[!h]
\begin{center}
\includegraphics[width=0.45\textwidth]{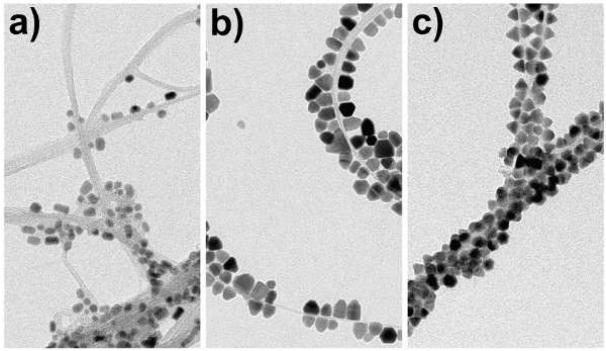}
\caption{\it CdSe NPs and CNTs obtained upon injection of 20~$\mu$L of HCl (20~$\mu$mol Cl) (a) after 3~h of reaction and (b) after 20~h. The injection of 10~$\mu$L of DCE (0.25~mmol Cl) instead yields the composites (c) after 20~h of reaction. Scale bars correspond to 100~nm.}
\end{center}
\end{figure}

Concerning the synthesis, we found that the injection of exclusively DCE without nanotubes and its subsequent removal under vacuum yields pyramidal-like nanoparticles. Thus, small traces of DCE are able to change the shape of the particles during the reaction. We studied some other parameters like the amount of CNTs. We observed that CNTs concentrations between 0.2-2~mg/run do not influence the shape of the particles or the CNT coverage. We also studied the effect of a second degassing process (supporting information) after Cd complexation in order to remove the released water according to the following the reaction~\cite{C13,C14}:

\begin{equation*}
	CdO + 2(ODPA) \rightarrow Cd(ODPA)_{2} + H_{2}0
\end{equation*}

Interestingly, the presence of water improves the coverage of the nanoparticles around the tubes once the particles are pyramidal-like. Several works have addressed the formation of anhydrides of ODPA as products of this reaction~\cite{C06,C15}. The presence of water would prevent the formation of such anhydrides, influencing the capping environment of the nanoparticles. Taking into account that both water and DCE influence the reaction, HCl might be formed. Previous work reported on the change of the chemical configuration of molecules upon sonication in chlorinated solvents, in particular DCE~\cite{C16}. Traces of HCl in commercially available Cl-containing solvents or formation of HCl upon sonochemical degradation may change the photophysical properties of different compounds. In order to prove the influence of HCl in the reaction, we performed the reaction by injecting CNTs in a Cl-free solvent like toluene. The addition of CNTs previously sonicated in a solvent is important to reduce the agglomeration of the CNTs or the CdSe-CNTs composites. Injecting CNTs in toluene (followed by its removal under vacuum) does not yield the formation of pyramidal-like particles (see next section). However, the injection of HCl or DCE to this reaction mixture after the formation of rod-like nanoparticles yield similar CdSe-CNT composites as the ones obtained when CNTs are injected dispersed in DCE, as can be seen in Fig.~2. To a mixture of Cd(ODPA)$_{2}$ in TOPO, CNTs in toluene were included in the reaction mixture before the Se precursor. After the nucleation and growth of rods (30~min after the Se injection), 20~$\mu$L of HCl aqueous solution 1N were added, without apparent formation of the CdSe-CNTs composites after 3~h, as shown in Fig.~2a. However, 17~h later, the samples obtained are similar to the ones obtained when CNTs are added dispersed in DCE (Fig.~2b). The amount of HCl added corresponds to 20~$\mu$mol of Cl. Due to the apparent degree of ripening less polydisperse particles could be obtained by adding a smaller volume of HCl. The same experiment performed by injecting 10~$\mu$L of DCE (0.2~mmol of Cl) after 30~min of reaction is shown in Fig.~2c. In this case, the nanoparticles are more monodisperse, what is related to fewer mols of Cl added. The sample was taken after 20~h of reaction, as in b).

\begin{figure}[!h]
\begin{center}
\includegraphics[width=0.45\textwidth]{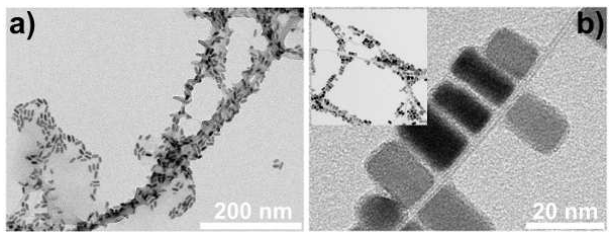}
\caption{\it (a) Nanotubes covered by the products of the reaction and CdSe rods after 30~min. (b) Composites obtained after 48~h of reaction when the CNTs where injected in toluene 24~h after nucleation. Inset: lower magnification.}
\end{center}
\end{figure}

Since the presence of water is non-avoidable even when degassing the Cd source after complexation, we propose that HCl may form when injecting CNTs dispersions in DCE in the reactant mixture containg Cd-ODPA, TOPO and water. In a quick experiment it could be seen how the pH value of DCE/water mixtures slightly decreases at around 80$^{\circ}$C, suggesting the possible formation of HCl under these conditions.

But what is the role of DCE or HCl during the reaction? Its presence in the reaction modifies the Cd-ODPA complex and/or destabilizes the ligand environment of the rod-like nanoparticles, boosting an intraparticle ripening that yields pyramidal-like ones. Together with an Ostwald ripening mechanism, the non-stable nanoparticles attach to the carbon lattices that act as new ligands. If formation of new Cd species and/or removal of ODPA ligands by protonation or nucleophilic displacement are the mechanisms of such destabilization is definitely an issue to be addressed by further studies. Covalent functionalization of the CNTs by the presence of the acid is excluded, as previous Raman characterization confirmed~\cite{C09}.

The requirements for the attachment of the nanoparticles to the tubes are (1) CNT surfaces free of ODPA and TOPO and (2) low passivated nanoparticle surfaces. Several examples clarify these two assessments:

\begin{figure}[!h]
\begin{center}
\includegraphics[width=0.45\textwidth]{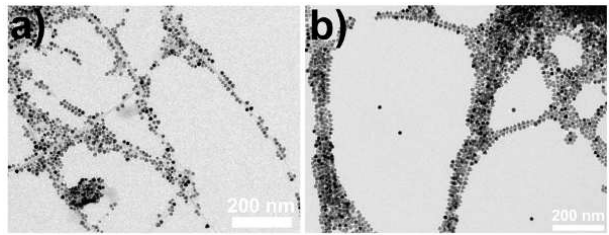}
\caption{\it (a) CNT covered with CdSe nanoparticles washed and incubated with CNT under stirring for 48~h. (b) Composite obtained after 48~h at 255$^{\circ}$C following the procedure described in Reference~\cite{C09}.}
\end{center}
\end{figure}

(1) \textbf{CNT surfaces free of ODPA and TOPO.} We observed that at early stages of the reaction (i.e. 30~min after nucleation), the rods are well capped (according to inter-particle distances measured by TEM) and the nanotubes are partially covered by the organic reactants (Fig.~3a), both facts preventing a successful attachment. At this stage, washing these samples in toluene does not completely remove the organic products around the CNTs. A similar organic coverage has been observed by the precipitation of polymers on CNTs by means of the so-called "antisolvent-induced polymer epitaxy method"~\cite{C17}. However, the coverage of the CNTs by the organic products is removed as the reaction evolves (i.e. 24~h after nucleation). The fact that at this later stage the CNTs do not seem to be wrapped by organics is related the evolution of the co-products in the reaction. To understand the role of the wrapped CNT surface, we performed experiments adding CNTs in toluene 24~h after the rods are formed. Fig.~3b shows the obtained samples 48~h after the CNTs injection.  

\begin{figure}[!h]
\begin{center}
\includegraphics[width=0.45\textwidth]{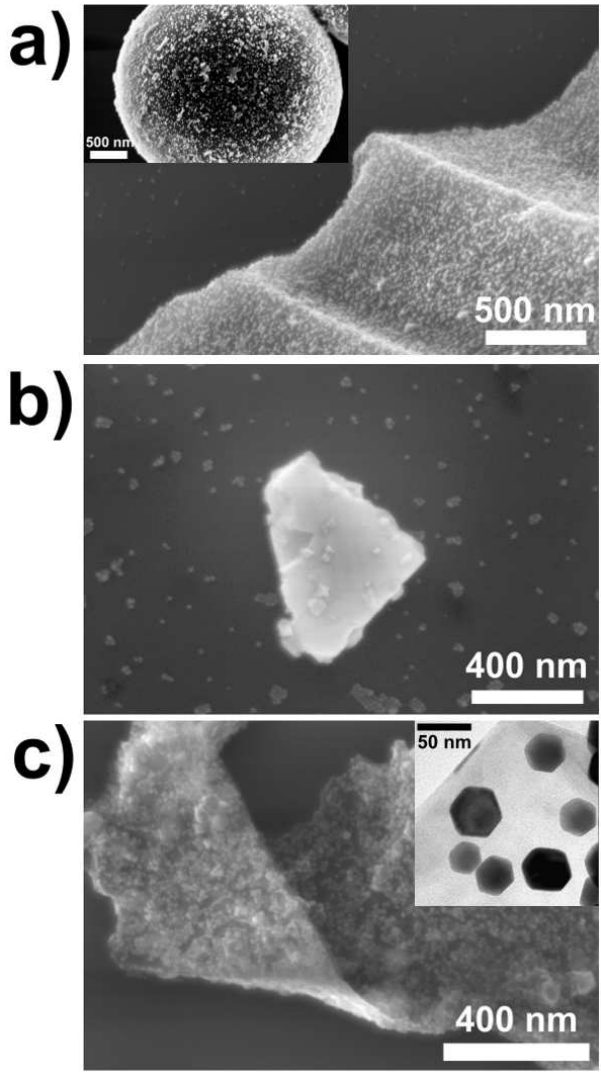}
\caption{\it Glassy carbon surface covered with CdSe nanoparticles (a) and diamond powder (b) where CdSe nanoparticles can be distinguished on the SiO$_{2}$ substrate. The inset in (a) shows a glassy carbon microsphere covered by CdSe nanoparticles. (c) Graphene layers covered by CdSe nanoparticles on a SiO$_{2}$ surface.}
\end{center}
\end{figure}

In this case, the CNTs surface is free from the above described wrapping products and the NPs also show a certain tendency to decorate the CNTs. The amount of nanoparticles covering the CNTs is substantially lower than in the case reported for pyramidal-like nanoparticles (Fig.~4b). The particles do not show prominent (10-1) facets, as for pyramidal-like ones, but mainly very flat (100) and (00-1). Thus, the pyramidal-like shape of the nanoparticles is clearly related to the effect of DCE or HCl and the attachment required, among other conditions, CNTs surfaces free of organics. This result also points out that CNTs stabilize flat facets of the particles. Nanoparticles showing non-defined and more rounded facets (like the ones shown in Fig.~2a) do not attach to the carbon lattice.

(2) \textbf{Low passivated nanoparticle surfaces.} In order to understand the role of the passivating ligand environment, we performed experiments involving washing rod-like CdSe nanoparticles and mixing them with CNTs under stirring at room temperature for 24~h (procedure B in the supporting information). The mixing was performed in toluene and in a destabilizing solvent like DCE. While in toluene the nanoparticle dispersion is stable, in DCE the particles preferentially decorate the CNTs, demonstrating again the ligating effect of CNTs. This result suggests the possibility to obtain to obtain comparable CdSe-CNTs composites at room temperature. In order to address this issue we mixed pyramidal-like nanoparticles obtained by the injection of DCE during the synthesis (without CNTs) with CNTs dispersed in toluene at room temperature (Fig.~4a). For comparison, Fig.~4b shows a sample obtained after 48~h, following the procedure described in the experimental section (A) at 255$^{\circ}$C. Although the amount of particles obtained is higher at higher temperature, a clear tendency of attachment is apparent for samples incubated at RT. Thus, low passivated (or eventually non-passivated) nanoparticles attach to the CNTs surface.

Finally, in order to study the role of the hybridization state and thus the crystal structure of carbon on the attachment, we performed experiments where CNTs were substituted by other allotropes of carbon such as glassy carbon or diamond powder. The former is known to possess a sp$^{2}$ lattice and the latter is sp$^{3}$ hybridized. The experiments were carried out as in the case of CNTs by injecting the carbon media dispersed in DCE. Fig.~5a shows a SEM picture of glassy carbon powder covered by CdSe nanoparticles. The inset shows glassy carbon (in form of spheres) covered by CdSe nanoparticles. TEM inspection of the nanoparticles not attached to the glassy carbon showed pyramidal shape. However, no apparent attachment is observed in the case of the diamond powder (Fig.~5b) what points to a clear role of the carbon crystalline sp$^{2}$ lattice, as has been already reported~\cite{C12}. Our results support recent work reporting a fullerene-like structure for commercial glassy carbon~\cite{C18}. Due to the increasing interest on the use and understanding of the physics of graphene, a commercial piece of HOPG was sonicated in DCE before injection. The sonication results in the formation of small flakes that may eventually contain a single layer of graphite produced by exfoliation, the so called graphene. Fig.~5c shows a SEM picture of several graphene layers covered with CdSe nanoparticles. The inset shows a TEM detail where the nanoparticles show the (001) facet connected to the carbon substrate. Although a clear assessment to a single monolayer cannot be done without further characterization, the results show that it is possible to attach CdSe nanoparticles to graphene by means of this approach, which we believe may bring fascinating physical and chemical properties to explore.

In conclusion, in this letter we give insights to the mechanism by which CdSe nanoparticles attach to CNTs following the hot injection method. It was observed that the presence of water improves the nanotube coverage while DCE, HCl or in general traces of a Cl are responsible for the shape transformation of the nanoparticles and further attachment to the carbon lattice. The experiments also show that the mechanism taking place involves the right balance of several factors, namely, low passivated nanoparticle surface, particles with well defined crystallographic facets, and interaction with an organics-free sp$^{2}$ lattice. Furthermore, this procedure can be extended to cover graphene by quantum dots.

\section*{Acknowledgment}
B.H Juárez thanks the European Comission for her Marie Curie Felowship.

\clearpage

\end{document}